\documentstyle[psfig]{article}
\begin{document}

\title{\sc A Modified  Parallel Tree Code for N-body Simulation  
of the Large Scale Structure of the Universe}

\author{U. {\bf Becciani}, V. {\bf Antonuccio-Delogu} and M. {\bf Gambera} \\
Osservatorio Astrofisico di Catania\\Via S. Sofia 78, I-95125 Catania - Italy \\ e-mail: ube@sunct.ct.astro.it}

\date{}
\maketitle

\begin{abstract}
N-body codes for performing simulations of the origin and evolution of 
the 
large scale structure of the universe have improved 
significantly 
over the past decade in terms of both  the resolution achieved and 
the reduction of the CPU time. However, state-of-the-art  N-body 
codes hardly allow one to deal with particle numbers larger than a few  $10^7$, 
even on the largest parallel systems. In order to allow simulations with larger 
resolution, we have first reconsidered the 
grouping strategy as described in J. Barnes (1990, {\it J. Comput. Phys.} {\bf 87}, 161) (hereafter B90) 
and applied it with some modifications to our WDSH-PT (Work and Data SHaring - 
Parallel Tree) code (U. Becciani et al., 1996, {\it Comput. Phys. Comm.} {\bf 99},1).   
In the first part of this paper we will give a short description of the
 code adopting the algorithm of J. E. Barnes and P. Hut (1986, {\it Nature}, {\bf 324}, 446)  and
in particular  the memory and work distribution strategy 
applied to describe  the {\it data distribution} on a CC-NUMA machine like the 
CRAY-T3E system. 
In very large simulations (typically $ N \geq 10^7$), due to network 
contention and
the formation of clusters of galaxies, an uneven load  easily verifies.
To remedy this, we have devised an automatic work redistribution 
mechanism which provided a good dynamic load balance without adding  
significant overhead. In the second part of the paper 
we describe the modification to the Barnes  grouping strategy we have devised 
to improve the performance of the WDSH-PT code. We will use the 
property that nearby particles  have similar interaction lists. This idea has 
been 
checked in B90, where an interaction list is built which applies everywhere 
within 
a cell $C_{group}$  containing a small number of particles $ N_{crit}$. B90 
reuses this 
interaction list for each particle $ p \in C_{group}$ in the cell in turn. 
We will assume each particle 
$ p$ to have the same interaction list. We consider that the agent force 
${\bf F}_p$ 
on a particle $ p$ can be decomposed into two terms 
${\bf F}_p = {\bf F}_{far} + {\bf F}_{near}$. The first term ${\bf F}_{far}$ is 
the same for each particle in the cell and  is generated by the 
interaction between a hypothetical particle placed in the center of mass 
of the $C_{group}$ and the farther cells contained in the interaction list. 
${\bf F}_{near}$ is different for each particle $p$ and  is 
generated by the interaction between $ p$ and the elements near $C_{group}$.
 Thus it has been possible to reduce the CPU time and increase the code performance. 
This
 enable us to run simulations with a large number 
of particles ($ N \sim 10^7 \div 10^9$) in nonprohibitive CPU times.

\end{abstract}

\section{INTRODUCTION}

N-body codes are one of the most important tools of 
theoretical cosmology \cite{ber91} because they offer the possibility of 
simulating most of the gravitational processes driving the formation 
of the  large scale structure of the universe (hereafter LSS) 
\cite{bar86}\cite{her87}\cite{dub88}. These simulations are often used to check 
cosmological 
models and to constrain the free parameters of these models which cannot be 
fixed 
either theoretically or observationally.\\
The typical mass scale for 
gravitational instability, the Jeans mass, has a value of $\approx 10^{6.5}$
solar masses (1 solar mass $\approx 1.9\times 10^{33} g$) at the recombination
epoch, and it gives the approximate size of the first objects forming by
gravitational collapse at that epoch. On the other hand, the largest structure
we see in our Universe today, the ``Supercluster'' of galaxies, has a mass
in excess of $\approx 10^{18}$ solar masses. Moreover, the gravitational force 
has a truly long-range character, which makes it impossible to introduce 
reasonable upper cutoffs in the mass range. For all these reasons,  one 
would like to be able to perform simulations 
spanning more than 12 orders of magnitude in mass, but present-day
state-of-the-art software and hardware technology does not allow 
simulations with more than $\approx 10^9$ bodies. For these reasons, 
the quest for increasingly efficient algorithms is still in progress.
However, the importance of making N-body simulations is 
clear to several authors \cite{dub96}\cite{kra97}\cite{kuh96}. 
During the past years N-body codes have been much improved and 
 applied successfully 
to various problems in galaxy dynamics, galaxy formation, and cosmological 
large structure formation. Nevertheless, the computational expense has 
remained prohibitive for $ N > 10^9$, even using tree-based algorithms 
on the most powerful computers.\\
The situation is even worse for other N-body algorithms. 
The N-body direct evolution method scales as $O(N^2)$, which  makes it
impossible
to run simulations with more than $10^4$ particles. To overcome this difficulty,
and when high accuracy is required, alternative numerical methods based on 
hierarchical force-computation algorithms are widely used. The recent effort 
 has  addressed  the production of new software 
and algorithms for the new  generation of high-performance computer systems. 
The ultimate target is  an implementation of
 the tree N-body algorithm to run simulations with higher  accuracy 
and particle number, decreasing the cost of the simulation in terms of CPU time
  and increasing  performance in terms of number of particles/second 
elaborated when
 running on MPP systems. \\
Among the tree algorithms designed to compute the gravitational force in N-body 
systems, one of the most used and powerful in modern cosmology is that by Barnes and Hut (BH) \cite{barh86}.  
The BH octal-tree recursive method is inherently adaptive and allows one to 
achieve 
a higher mass resolution even if parallel implementation of
this algorithm \cite{bec96}\cite{sal90}  suffers from a serious drawback: it can 
easily run into 
imbalance as soon as the configuration evolves, causing  performance 
degradation. 
In this paper we present a modified version of the BH algorithm in which we have
introduced an enhanced 
grouping strategy. We will show how this feature allows an increase in 
performance
 when we consider N-body  simulation with a large number of particles ($ N \geq 
10^6$). The code we present incorporates  fully periodic boundary conditions using the Ewald method, without the use of fast Fourier transform techniques \cite{her91}.\\ 
In Section 2  we give a brief description of 
our N-body  
parallel code, based  on the BH tree algorithm,
and the dynamic load balance (DLB) policy adopted. In Section 3 we 
describe our enhanced grouping strategy.  In Section 4 we 
show the results of our tests and in Section 5 we report our conclusions.

\section{THE PARALLEL CODE}

Since the publication of the monograph by Hockney and Eastwood in 1981 \cite{hoc81}, a new class of 
particle simulation methods \cite{app85}\cite{barh86}\cite{jer85}\cite{por85}\cite{pre86} has emerged as an alternative to particle-particle 
(PP) \cite{aar71}\cite{hen64}\cite{von60}  particle-mesh (PM) 
(for a review of this method see \cite{bir85}) and 
particle-particle-particle-mesh ($P^{3}M$) \cite{eas74} methods. 
These new methods are characterized by  the particles being arranged
into a hierarchy of clusters, which span the full range of length scales 
from the minimum interparticle spacing up to the diameter of the entire 
system. These methods are usually known as {\it tree methods} or 
{\it tree codes} because of the data structures used.
With these new methods, the short-range force on a particle $ p$ is calculated 
as a direct sum over nearby particles. Remote bodies are organized into 
 groups which become progressively larger with the distance from the 
particle;  
then a multipole expansion of the potential of each cluster about its 
center of mass is performed. The long-range contribution to the 
acceleration is given by the sum of the particle-cluster interactions.

\subsection{The Barnes-Hut Tree Algorithm}

The BH algorithm works using a hierarchy of cubes arranged in an octal-tree
 structure; that is, each node in the tree has eight siblings and each node represents a physical volume of the space. The total mass of all particles 
within a given volume and their centers of mass are stored at the 
corresponding node. Thus, the system is first surrounded by a single cell 
(cube) encompassing all  the particles. This main cell (called {\it root}) is
subdivided into eight subcells of equal volume, each containing 
its own subset of particles.
 Each subcell in turn is subdivided into 
eight new subcells and so on. This procedure is repeated until each 
cell at the lowest level contains only one  particle.\\
 . The force on any given $ p$ is then the 
sum of the forces by the nearby particles plus the force by the distant cells
whose mass distributions are approximated by multipole series truncated 
typically at the quadrupole order \cite{her87}. The criterion for 
determining whether a cell 
is sufficiently distant for a multipole force evaluation (that is, for 
approximating the cell as a multipole) is based on an opening angle 
parameter $ \theta$  given by
\begin{equation}
\frac{C_l}{d} \le \theta ,
\label{eq:un}
\end{equation}
 where $C_l$ is the size of the cell and $d$ is the distance of 
$ p$ from the center of mass of the cell. 
Smaller values of $ \theta$ lead to more cell opening and more accurate 
forces (for $ \theta = 1$ we have an error lower than 1\% on the 
accelerations \cite{her87}). The equations of the dynamics are solved using
 the Leapfrog integrator.\\

\begin{figure}
\begin{center}
\begin{minipage}{12cm}
\centerline{\psfig{file=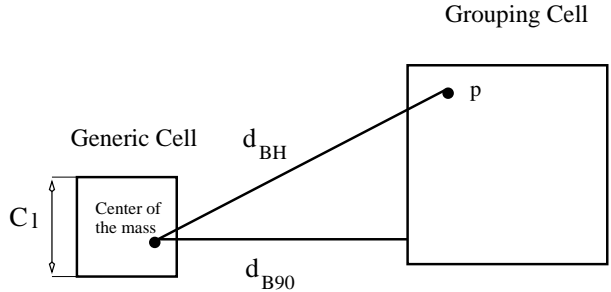,width=8cm}}
\caption[h]{\small Interaction list formation in BH code.}
\end{minipage}
\end{center}
\end{figure}

\subsection{Data Distribution and DLB}

In our parallel implementation of the BH tree algorithm, 
using the  PGHPF/CRAFT (an implementation of High Performance Fortran by the 
Portland Group) programming environment  for the Cray T3E system, 
we have exploited both the {\it Data Sharing} and the {\it Work 
Sharing} programming models. The flexibility of the PGHPF/CRAFT environment 
allows one to mix these two modes in order to gain the maximum efficiency 
and speed-up. We can distinguish two main phases in our code structure: the Tree\_formation (TF) and  the Force\_compute (FC). A data distribution in  contiguous blocks 

\vspace*{0.3cm}

\noindent \hspace*{1.5cm} !HPF\$ DISTRIBUTE PARTICLE\_ATTRIBUTE(BLOCK,*)
\vspace*{0.3cm}

\noindent and alternatively, a fine grain distribution

\vspace*{0.3cm}
\noindent \hspace*{1.5cm} !HPF\$ DISTRIBUTE TREE\_ATTRIBUTE(CYCLIC,*)
\vspace*{0.3cm}

\noindent were adopted to distribute the particle data properties and  the tree data properties.\\
The !HPF\$ DISTRIBUTE  directive of the PGHPF/CRAFT compiler allows us to consider an array like  PARTICLE\_ATTRIBUTE (or TREE\_ATTRIBUTE) as a unique large array, accessible from all the processors,
 the array being physically distributed in the local memory of all the processors.
We used two different sets of initial conditions, 
namely uniform and clustered  distributions
having 2 million particles each, and they were carried out using from 16 to 
128 PEs.
Our results show that the higher code performances are obtained using a fine grain tree data distribution and a coarse grain bodies data distribution.
A detailed description can be found in \cite{bec96}.
The static array distribution, fixed as described above, allows each PE to
cooperate during the TF  phase by using principally the DO INDEPENDENT structure
of PGHPF that is a synchronous mechanism, and then to execute the FC phase in
asynchronous mode. To minimize the communication overhead, each PE executes the FC phase
mainly on the local residing bodies. The BLOCK distribution arranges bodies with 
the nearest logical  number (near in the space) in the same PE local memory, or in the nearest PEs.
Using the above mentioned data distribution, each PE has a block of closed
 bodies in the 
local memory ($Np=N_{bod}/N\$PEs$, where  N\$PEs is the number of processors
used for the simulation);in an initial condition with a uniform distribution, the PEs having 
extreme numeration in the 
pool of available PEs have a lower load at each time-step. 
The load imbalance is enhanced when a clustered situation  occurs during
the system evolution. The PEs having bodies in clustered regions have a greater 
workload since the load of the FC phase increases as the mass density grows.
The technique we follow to perform a load redistribution
among the PEs is
to assign  each PE to execute this phase {\it only} for a fixed 
portion of the bodies residing in the
local memory $NB_{lp}$ given by\\
\begin{equation}
NB_{lp}=(N_{bod}/N\$PEs)P_{lp},
\label{eq:dlb1}
\end{equation}
where $P_{lp}=$ const. ($0 \leq P_{lp} \leq 1$).

The FC phase for all the remaining bodies\\
\begin{equation}
N_f=N\$PEs(N_{bod}/N\$PEs)(1 - P_{lp})
\label{eq:dlb2}
\end{equation}
is executed by all the PEs that have concluded the FC phase 
for the assigned $NB_{lp}$ bodies. No correlation is considered between the 
PE memory 
location of the body belonging to the  $N_f$ set and the PE that computes
the FC phase on it.
The  results imply that it is possible to fix a $P_{lp}$ value that allows the best code performances. Data already presented in \cite{bec97}
show that  
it is convenient to fix the $P_{lp}$ value near  $0.25$,
which is the value that  maximizes the load balance for N-body simulations of the LSS
 both in uniform and clustered situation.
\begin{figure}
\begin{center}
\begin{minipage}{12cm}
\centerline{\psfig{file=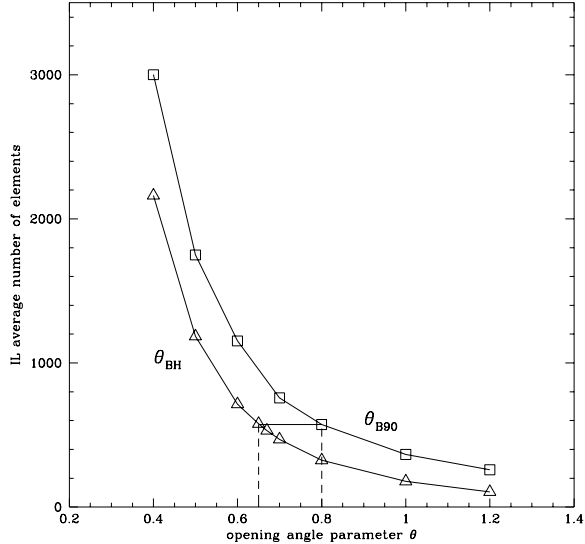,width=8cm}}
\caption[h]{\small Interaction list average length for BH and B90 algorithms.}
\end{minipage}
\end{center}
\end{figure}

\section{GROUPING}

Our work and data sharing-parallel tree (WDSH-PT)  code 
is principally aimed at running LSS  cosmological simulations with
a number of particles as high as possible using supercomputers such as Cray 
T3E systems. In order to increase the code efficiency, we adopt initially
the grouping method proposed by Barnes \cite{bar90} and introduce a modified implementation of his grouping policy
yielding very high gains
in the code performances with the same accuracy.

\subsection{B90 Grouping}

To compute the force on a body, the BH algorithm needs to build an 
interaction list ($IL$) for each particle $p$. Starting from the root cell, a tree inspection is done and the opening angle 
parameter $\theta$ is used  to evaluate whether a cell must be opened or
 closed as mentioned above. If a cell has dimension $C_l$ and distance $d$ 
from the particle $p$ 
so that  Eq. (\ref{eq:un}) is verified, the cell is closed, it is added to 
the $IL$, and its subcells are not investigated further. 
Otherwise the cell is opened and its subcells are investigated in the same way. Bodies belonging to an opened cell are  added to the $IL$.\\
Next, the force on the body is
 computed using the monopole and  
quadrupole momenta for all the cells  in the list.\\

\paragraph{BH timing} The tree inspection phase represents a sizeable task to compute the force 
because the cell opening criterion is applied many times for each particle. The CPU time $T_o$ to compute the force in a  time-step for all the $N$ particles is

\begin{equation}
T_o = N \langle T_{l} \rangle + N \langle T_{f} \rangle ,
\label{eq:tre}
\end{equation}

\noindent where $\langle T_{l} \rangle$ is the average time to build an $IL$ and $\langle T_f \rangle$ is the average time
to compute the force on each particle using the interaction list.\\
\begin{figure}
\begin{center}
\begin{minipage}{12cm}
\centerline{\psfig{file=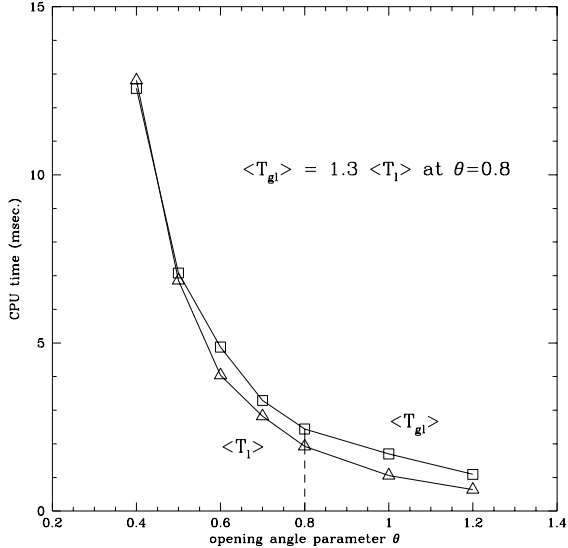,width=8cm}}
\caption[h]{\small Measured $\langle T_l \rangle$ and $\langle T_{gl} \rangle$ using WDSH\-PT code in a Cray-T3E 1200 system.}
\end{minipage}
\end{center}
\end{figure}

\paragraph{B90 timing} The basic idea of  B90 was to build a unique interaction list
that allows the force for a {\it group} of particles inside a region;
i.e., a cell $C_{group}$ of the tree {\it (grouping cell)}, to be computed reducing the number of  tree inspections to build the $ILs$. 
B90 builds an IL that applies
everywhere within $C_{group}$  
and reuses this IL for each particle $p \in C_{group}$  in turn.
 In this way it is possible to reduce  the 
tree inspection phase. The CPU time $T_g$ for B90 may be written as 

\begin{equation}
T_g = N_{gc} \langle T_{gl} \rangle + N \langle T_{gf} \rangle ,
\label{eq:du}
\end{equation}

\noindent where 
\begin{itemize}
\item{$ N_{gc}$ is the number of grouping cells (assuming that each body is
inside a group region);}
\item{$\langle T_{gl} \rangle$ is the 
average time to build an interaction list for a group;}
\item{ $\langle T_{gf} \rangle$ is the 
average time to compute the force on a particle using the list formed for
the group.}
\end{itemize}

In the following  paragraphs we will compare the $T_g$  time with the $T_o$ time considering 
the generic case  $\theta=0.8$ .  We notice that different values of $\theta$ give similar results, as shown by the accompanying figures.

\paragraph{B90 opening criterion} The original BH
 algorithm adopts an opening criterion $\theta_{BH}$, based on the distance between
the position of the $p$ particle and the center of mass of the remote cells, the $IL$ length ($L_{IL}$)
 being proportional to ${(\theta_{BH}^3)}^{-1}  logN$. In order to have the same accuracy as the original algorithm, the interaction list of the  grouping cell is formed using Eq. (\ref{eq:un}) but
now the $d$ term is computed in terms of the distance from the center of mass of an
 inspected cell and the edge of the   grouping cell, as shown in Fig. 1 ($d_{B90}$ is used instead of $d_{BH}$).

This implies that the $IL$ formed using the  grouping cell
contains more elements than the $IL$ formed by applying 
the original BH algorithm. \\

\paragraph{B90 interaction list increment} The B90
adopts an opening criterion  ($\theta_{B90}$) based on the distance between
the edge of $C_{group}$ and the center of mass of the remote cells. In this
case the $IL_g$ length will be proportional to $ {\theta_{B90}^3}^{-1}  logN$.
Moreover, the B90 criterion uses $\theta_{B90}$ numerically equal to $\theta_{BH}$ when using the original BH algorithm. We found experimentally the relation between $\theta_{B90}$ and $\theta_{BH}$, using 2 million particles in a uniform distribution (see Fig. 2): this relation agrees with data in Salmon \cite{sal90}. 
A typical value used as opening criterion to run simulations  for the 
LSS is $\theta_{BH}=0.8$. Consequently
 we consider $\theta_{B90}=0.8$ , which, in terms of $IL$ length increment, corresponds to
running a simulation with  $\theta_{BH}=0.6$.\\

\paragraph{B90 timing vs BH timing} Figs. 3 and 4  show the relationship $\langle T_{gl} \rangle - \langle T_l \rangle$ and $\langle T_{gf} \rangle -  \langle T_f \rangle$
with $\theta$ ranging between $0.4$ and $1.2$. Considering $\langle T_{gl} \rangle = 1.3  \langle T_l \rangle$   and 
$\langle T_{gf} \rangle = 2.2 \langle T_f \rangle$  with $\theta = 0.8$,  Eq. (\ref{eq:du}) may be rewritten using the above  relations as follows:

\begin{equation}
T_{g} = 1.3 \frac {N} {\langle N_{gp} \rangle}   \langle T_{l} \rangle + 2.2 N   \langle T_{f} \rangle ,
\label{eq:Tgl1}
\end{equation}

\begin{figure}
\begin{center}
\begin{minipage}{12cm}
\centerline{\psfig{file=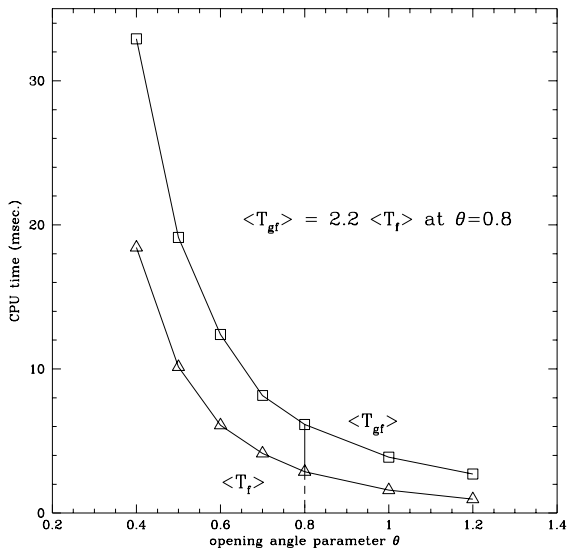,width=8cm}}
\caption[h]{\small  Measured $\langle T_f \rangle$ and $\langle T_{gf} \rangle$ using WDSH\-PT code in a Cray-T3E 1200 system.}
\end{minipage}
\end{center}
\end{figure}

\noindent where $\frac{N}{\langle N_{gp} \rangle} = N_{gc}$, and $\langle N_{gp} \rangle$ is the average number of particles in a grouping cell. \\
For a large number of systems and in particular for our WDSH-PT code running
on the T3E system, $\langle T_f \rangle$ ranges from $ \sim 1.2 \langle T_l \rangle$ to  $ \sim 1.5 \langle T_l \rangle$ at $\theta=0.8$. Considering these figures, we obtain from Eqs. (\ref{eq:tre}) and (\ref{eq:Tgl1}), respectively,

\begin{equation}
T_o = 2.2 N \langle T_l \rangle
\label{eq:To_m}
\end{equation}

and

\begin{equation}
T_g = 2.2 N \langle T_l \rangle (\frac {1.3}{2.2 \langle N_{gp} \rangle} +  1.2)
\label{eq:TgTo}
\end{equation}
  
\noindent and then  $T_g > T_o$.\\

A real gain could be obtained using B90 if the CPU time spent to form the interaction list
 were longer than the phase to compute the force on the
particle. The results reported in \cite{bar90} demonstrate that if $\langle T_{l} \rangle \gg \langle T_{f} \rangle$ the code 
performance is  
between two and three times faster of the BH algorithm, and a 
good choice for the $N_{crit}$ value is about 32 \cite{dub96}.

\subsection{The Modified Grouping Strategy for LSS Simulations}

We will now describe the  modification we introduce in the 1999 version of our WDSH\-PT
 code  (WD99) 
to increase  performance even if $\langle T_l \rangle \leq  \langle T_f \rangle$.   The basic idea is 
to assign the same  $IL_g$ to each particle 
within a cell $C_{group}$, containing a maximum of $ N_{crit}$ particles.\\ 
We will not use the B90 criterion to build the interaction list. Instead, we will use 
the same $\theta_{BH}$ criterion used in the original BH algorithm. This criterion is applied to a hypothetical particle 
placed in the center of mass of the $C_{group}$, hereafter VB (Virtual Body) (Fig. 5). Moreover, we consider the
 $IL_g$ as formed by 
two parts given by
\begin{equation} 
IL_g = IL_{far} + IL_{near}
\label{eq:il_g}
\end{equation}
$ IL_{far}$ and $IL_{near}$ being two subsets of the interaction list. An element is included in one of the two subsets, using the following Sphere criterion for all the elements that satisfy Eq. (\ref{eq:un}).

\vspace*{1.0cm}

\noindent \hspace*{1.0cm}{\bf Define} $Sphere_{radius} = 3  \frac {Cellsize(C_{group}) \sqrt{3}}{2}$\\
\\
\\
\vspace*{0.3cm}
\hspace*{1.0cm}{\bf If} $Distance(IL_g(element),VB) >  Sphere_{radius}$ \\
\vspace*{0.3cm}
\hspace*{2.0cm} {\it Add element to} $IL_{far}$\\
\vspace*{0.3cm}
\hspace*{1.0cm}{\bf Else}\\
\vspace*{0.3cm}
\hspace*{2.0cm}   {\it Add element to} $IL_{near}$\\ 
\vspace*{0.3cm}
\hspace*{1.0cm}{\bf Endif}\\
\vspace*{1.0cm}

\begin{figure}
\begin{center}
\begin{minipage}{12cm}
\centerline{\psfig{file=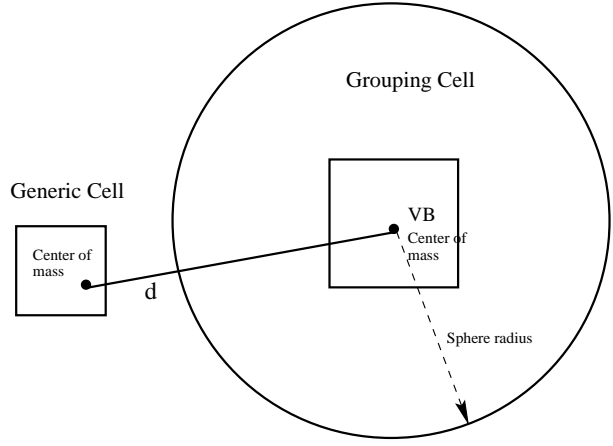,width=8cm}}
\caption[h]{\small $IL_g$ formation using the Sphere criterion.}
\end{minipage}
\end{center}
\end{figure}

\noindent Moreover all $p \in C_{group}$ are included in $IL_{near}$.

Using the two subsets it is possible to compute the  force ${\bf F}_p$ 
on a particle $ p \in C_{group}$ as the sum of two components,
\begin{equation} 
{\bf F}_p ={\bf F}_{far} +{\bf F}_{near} ,
\label{eq:F_p}
\end{equation}

\noindent where ${\bf F}_{far}$ is a  force component due to 
the elements listed in $IL_{far}$ and ${\bf F}_{near}$ is the force component
due to the elements in $IL_{near}$. 
We assume the  component  ${\bf F}_{far}$  
to be the same for each particle $p \in C_{group}$  and compute it considering the gravitational interaction between the VB
  and only the elements listed in 
$IL_{far}$, 
while the  ${\bf F}_{near}$ component is computed separately for each $p$ 
particle  by the direct interaction with the elements listed in 
$IL_{near}$.

Moreover, ${\bf F}_{near}$ contains a restricted
number of elements in comparison with the ${\bf F}_{far}$  list,
so we expect  a net gain in performance even if $ T_{l} \le T_{f}$. 
The gain that  is possible  depends on several parameters ($N_{crit}$, the size of the $C_{group}$ and 
 the {\it Sphere radius}), whose ranges of variation are constrained by the maximum allowed value of the overall error of the method, as we will describe in the following sections.

\subsection{Errors Analysis and Performance considerations}

Before showing the performance of our WD99 procedure in an N-body simulation of the large scale
structure of the universe, it is important that we perform an error analysis 
of the procedure itself.
Considering that the cumulative error, when simulations for the LSS studies are run using the original BH algorithm, is lower than $1\%$  \cite{barh89}, fixing the opening criterion
 $\theta=1$,  we will give some constraint concerning the
size of $C_{group}$ , the $N_{crit}$ value, and the {\it Sphere radius} needed to have  negligible
 cumulative error. The following sub-sections discuss  the
two main sources of error. 

\subsubsection{The Differences in the Interaction List} 

The first error source is that WD99  uses the  Sphere criterion and the VB to
create an
 interaction list $IL_g \equiv IL_{VB}$ and WD99 applies the
$IL_{VB}$ to all bodies $p \in C_{group}$. This approximation could introduce an
 error in the force value on the $p$ particle if the $IL_p$, created using the original BH algorithm, and the $IL_{VB}$ have a difference in the
 elements greater then $1\%$. As we found with our tests, in order to decrease this difference it is
necessary to limit  the size of $C_{group}$ that is equivalent to fixing a {\it critical
level} of the tree structure: cells above the critical level cannot form a grouping cell. The user has to fix the critical level considering
the density of local bodies  in the box where bodies are arranged: the critical level must be chosen in such a way as to make the difference between $IL_p$ and $IL_{VB}$ negligible (no more than $1\%$ of the elements).
It seems reasonable for a LSS simulation in a 50 Mpc box  with more than 2
million  particles to fix the critical level
as the sixth level of the tree. The next section shows in detail the obtained results. In any case, the 
cumulative error is very small considering the increase in accuracy compared with the original BH algorithm, due to the inclusion of all  $p \in C_{group}$ in the interaction list.

\subsubsection{Approximation of the Force Component}

The second error source is due to the assignment of ${\bf F}_{far}$, computed for the VB, 
to each $p \in C_{group}$. We found that the Sphere criterion allows
 us to reduce this error to values much lower than 0.01\%
 for $ N \ge 10^6$ if the dimension of
the $C_{group}$ cell is fixed with the  critical level as mentioned above, and
the {\it Sphere radius} is  three times the radius of the sphere 
enclosing the $C_{group}$ cell (Fig. 5).\\
Another important constraint to be fixed is the value of $N_{crit}$. All the elements
 $p \in C_{group}$ are listed in
the $IL_{near}$ list and there is a direct body-body interaction among
the $N_{gp}$  ($N_{gp} \le N_{crit}$) elements forming the group. This
introduces a term $O(N_{gp} N_p)$ in the algorithm complexity. In order to avoid
a decrease of the code efficiency 
and to maintain a good code accuracy, as with the original BH algorithm,
 it seems reasonable, running LSS simulation with more than 1 million particles,  to maintain $N_{crit} \le 32$.
We adopted, in our runs, a safe value  $N_{crit} = 16$ even if we obtained
good results with  $N_{crit} = 32$.\\
In the next section we show the errors obtained using the above-mentioned constraints when applying WD99
to LSS simulation with both uniform and clustered 
distributions.

\section{TESTS AND RESULTS}

We carried out many tests to estimate the error  introduced in the WD99 and obtained increased performances 
 using several values of 
$N_{crit}$. Therefore, this section is subdivided as follows: first we test whether our algorithm 
increases the average length of the interaction list, then we
measure the resulting percentage error,  and we conclude with an overall 
performance analysis.
As a test case we ran a simulation using  2 million particles for LSS 
in  a cubic region of 50 Mpc, starting from a homogeneous initial condition 
(redshift $Z = 50$) and reaching a clustered configuration (redshift $Z = 0$). We used  an opening angle parameter $\theta$ ranging
from  $0.8$  to $1.2$. Our tests
were executed on a Cray T3E system and the results  will be shown
 in the following sections.

\subsection{Measuring the Interaction List Length}

The aim of this first test is to verify that the WD99 algorithm
does not introduce a significant computational cost when 
 the force for a generic particle is computed. This measurement is
substantially performed on the average length of the $IL$ we form
adopting our code. 
Fig. 6 reports the result we obtain when the simulation evolves at
redshift $Z = 50$. 
\begin{figure}
\begin{center}
\begin{minipage}{12cm}
\centerline{\psfig{file=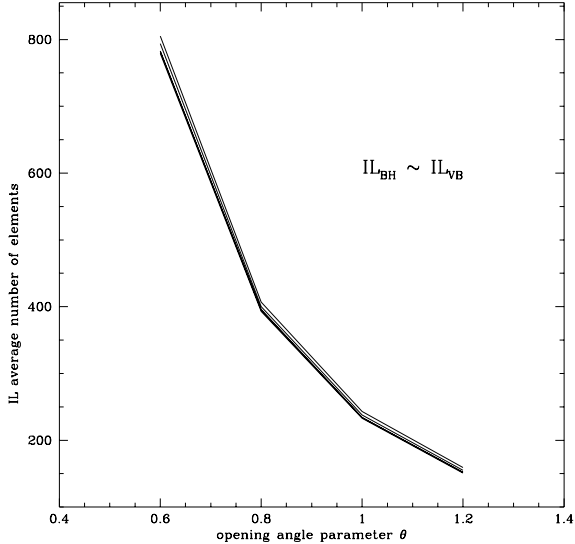,width=8cm}}
\caption[h]{\small Measurement of the average Interaction List length running the original BH code and our WD99 using critical level from 5 to 8.}
\end{minipage}
\end{center}
\end{figure}

Tests were executed for several values of redshift, but the differences between
BH  and our algorithm was computed only at the end of the run. The curves
were obtained by fixing  $N_{crit} = 32$ and varying the critical level 
from 5 to 8. In all cases the differences we obtained are negligible,
which means that the computed $IL$ for the VB (with the adopted Sphere criterion) is about equal to the $IL$ we obtain for a generic particle with
 the original BH algorithm. 
The first important result is that WD99 does not produce any increment in the IL length and
consequently $\langle T_l \rangle =  \langle T_{gl} \rangle$.

\subsection{Error Measurement}

We carry out this measurement in two phases. First we run a single time-step
of the 2-million-particle simulation at  redshift $Z = 50$.  
 We compare  the values we obtain running the BH original algorithm and 
the WD99.  As a reference case, we adopt the critical level equal to 6.

A similar comparison is made at $Z=0$ and the BH and WD99 histograms of the forces of each 
component are compared. The comparison  shows a negligible difference in 
the force distribution in a single time-step, at least an order of magnitude less than the error  of the original BH algorithm.\\
The second measurement is made analysing an entire system evolution. We start with the initial condition of 2 million particles 
with redshift $Z = 50$, $\Delta t = 0.001$, $\theta = 0.8$, and particle mass about $1.655 \cdot 10^{10}$ solar masses. The system evolution is carried out up to redshift $Z = 0$. The evolution
is executed with the original BH algorithm and with the WD99 code. As  reference case, we adopt the critical level equal to 6.
We measure the absolute error $\epsilon$ in the position of particles and in the velocities of particles in the mean square sense,

\begin{equation}
	\epsilon_{pos} = \frac {1}{N} \sqrt{\sum_{i=1}^{N}(X_{BH}-X_{WD99})^2+(Y_{BH}-Y_{WD99})^2+(Z_{BH}-Z_{WD99})^2}
\label{eq:eppo}
\end{equation}

and

\begin{equation}
	\epsilon_{vel} = \frac {1}{N} \sqrt{\sum_{i=1}^{N}(Vx_{BH}-Vx_{WD99})^2+(Vy_{BH}-Vy_{WD99})^2+(Vz_{BH}-Vz_{WD99})^2} .
\label{eq:epve}
\end{equation}

The study of the final evolution is  described in the next  sub-section. Here
we give the measured value at the end of the simulation:
$\epsilon_{pos} = 0.003$ and $\epsilon_{vel} = 0.01$.

Similar values are measured when running simulations with more than 2 million 
particles and with a critical level equal to 6. The obtained $\epsilon$ values
 lead us to conclude that the WD99 procedure does not introduce significant errors 
in comparison with the BH algorithm.

\subsubsection{Simulation Analysis}

The final stages  obtained running  simulations with the BH algorithm and 
the WD99 
code are very similar.
The two-point correlation function is defined as 

\begin{equation}
	\xi(r) = \frac{N_q}{\langle \rho \rangle N_c V} - 1 ,
\end{equation}

\noindent$N_q$ being  the number of pairs of particles with separations between $r$ and 
$r + \Delta r$, $V$ the volume considered, $N_c$ the particle number taken as centres, and $\langle \rho \rangle$ the mean particle density.
We calculate this function at redshift $Z=0$ for WD99 and BH algorithms. The values we obtain  are perfectly equal, and
the substructures we form (number and size) are identical.

\subsection{WD99 Performances}   

To conclude our WD99 description, we report the performances measured for the WD99 code (Fig. 7)
(including the boundary periodic conditions using the Ewald method \cite{her91}) and the performances of the 
original BH algorithm.
\begin{figure}
\begin{center}
\begin{minipage}{12cm}
\centerline{\psfig{file=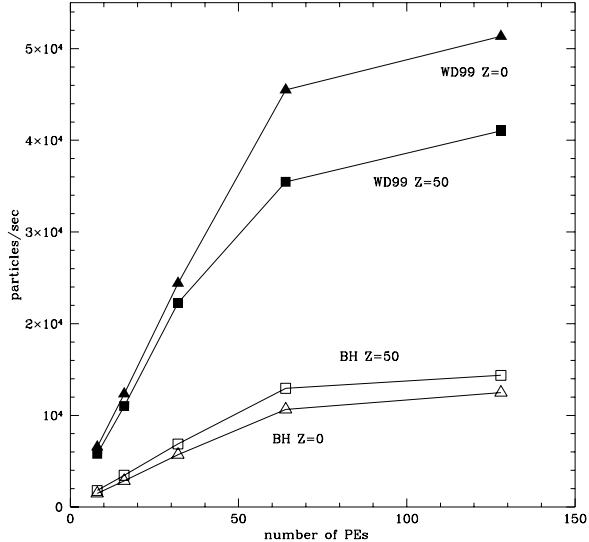,width=8cm}}
\caption{\small  A 2 Million of particles at $\theta =0.8$: BH algorithm and WD99 procedure performances. A critical level equal to 7 is fixed, and the measurement are
performed  for uniform ($Z=50$) and clustered ($Z=0$) system conditions.
The y scale measure the number of particle/sec we compute.}
\end{minipage}
\end{center}
\end{figure}
The measured  performances lead us to the  conclusion that when the system evolution is clustered ($Z=0$) the WD99 does not decrease the performance as the
BH algorithm. This effect is due to the nature of the WD99 algorithm, which has
a structure that increases the efficiency when clusters of particles 
are well closed. This important effect  allows
us to run simulations with very clustered systems, obtaining very good performance and negligible errors. 
Moreover, the efficiency of the WD99  increases by a factor of up
to  five at the redshift $Z=0$. The gain is enhanced when
 bigger simulations are run: a recent  simulation with 16 million  particles
 performed on the Cray T3E system  using WD99 showed an  increase in performance by a factor of 7 at the redshift $Z=0$.
We note that the gain obtained, in comparison with that obtained by the
original BH algorithm, is greater using  a lower critical level (5 or 6).
The gain is incremented  using a Sphere criterion with {\it Sphere radius} lower than the value we consider, having only a small increment in the global error.

\section{CONCLUSIONS AND FUTURE}

The code  WD99  is mainly used for LSS studies, but it could be
tested and used for other applications where  accuracy not higher  than 1\% is
necessary. Considering the high performances we obtained, 
the WD99 method may be  very successfully  applied  when clustered configurations such as
 galaxies or clusters of galaxies have to be studied.
The new approach  could be applied also to other fields of 
physics where collisionless systems are to be simulated, as in plasma and hydrodynamic studies.\\
The code is written in Fortran 90 with PGHPF/CRAFT, but the latest
 version (written in F90 and C languages) uses the one-side communication library SHMEM, allowing it to 
run on the ORIGIN 2000 systems. A new version will be implemented using
dynamical array allocation, and we are studying the implementation of the
parallel out-of-core \cite{sal97}, moving data in the disk. This version will
be developed for a CC-NUMA machine with MPI-2. We plan to have a freely available version of WD99 in 
October 2000. 

\section{ACKNOWLEDGMENTS}

All the tests were carried out using the CRAY T3E 1200/256 machine at the CINECA (Casalecchio di Reno (BO), Italy), a 256 PE system, using the financial support
 of the Italian Consortium CNAA (Consorzio Nazionale per l'Astronomia e l'Astrofisica). 
We  thank Dr. G. Erbacci of CINECA and Dr. A. F. Lanza of Catania Astrophysical Observatory for 
their useful help.

\end{document}